\documentclass[a4paper,11pt]{article}
\pdfoutput=1 

\usepackage{jinstpub} 
\usepackage{graphicx}
\usepackage{subcaption}
\usepackage{booktabs}
\usepackage{siunitx}
\usepackage{tabularx}
\usepackage{xcolor}
\usepackage{dcolumn}
\usepackage{multirow}

\newcommand{\ra}[1]{\renewcommand{\arraystretch}{#1}}
\newcommand{\tls}{\addlinespace[.6em]}

\title{A 4$\pi$ Fluorescence Detection Region for Collinear Laser Spectroscopy}

\author[a,1]{B. Maa\ss,\note{Corresponding author.}}
\author[a,b]{K. K\"onig,}
\author[a,c]{J. Kr\"amer}
\author[b,d,2]{A. J. Miller, \note{Current address: Sandia National Laboratories, Albuquerque, NM 87123, USA}}
\author[b,d]{K. Minamisono}
\author[a,c]{W. N\"ortersh\"auser}
\author[a]{F. Sommer}

\affiliation[a]{Institut für Kernphysik, Technische Universit\"at Darmstadt, 64289 Darmstadt, Germany}

\affiliation[b]{National Superconducting Cyclotron Laboratory, Michigan State University, East Lansing, MI 48824, USA}

\affiliation[c]{Helmholtz Research Academy Hesse for FAIR, Technische Universit\"at Darmstadt, 64289 Darmstadt, Germany}

\affiliation[d]{ Department of Physics and Astronomy, Michigan
State University, East Lansing, MI 48824, USA}

\emailAdd{bmaass@ikp.tu-darmstadt.de}

\abstract{
We report on a novel detection system for collinear laser spectroscopy which provides an almost 4$\pi$ solid angle for fluorescence photon detection by employing curved surface mirrors. Additional parabolic angular filters offer passive stray light suppression and can be configured to match the experimental conditions. The mirror surfaces have an excellent reflectivity over a broad band of wavelengths in the optical spectrum and can be substituted to expand the wavelength acceptance range even further. Experiments with this system were performed at two collinear laser spectroscopy setups, including the laser spectroscopic investigation of $^{36}$Ca using rates of 25\,/s at NSCL, MSU.
}

\begin{document}
\maketitle
\flushbottom

\section{Introduction}
\begin{figure}[ht]
    \centering
        \includegraphics[width=0.98\textwidth]{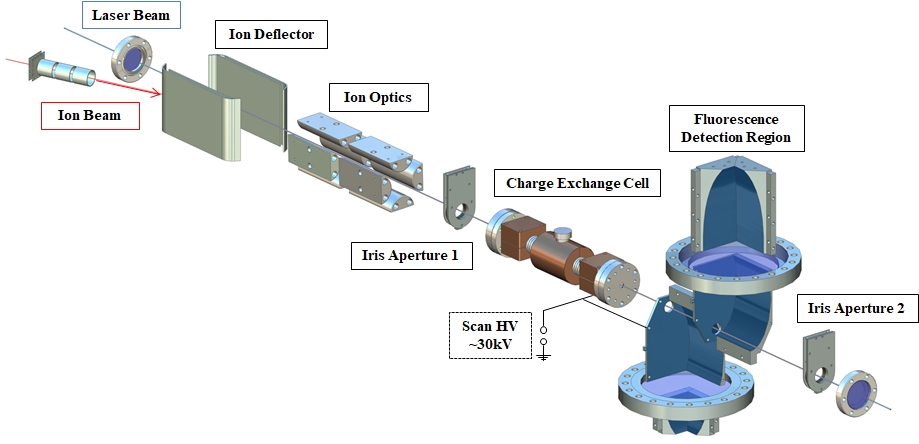}
    \caption{In a typical setup for collinear or anti-collinear laser spectroscopy, the ions of interest are deflected into a linear vacuum beam line with the help of ion optics. In the linear section, ions are overlapped with a laser beam using a set of apertures. A charge exchange cell is used to neutralizing the beam in-flight when spectroscopy on atoms is performed. In the fluorescence detection region, photons emitted from the beam are collected when the investigated particles are in resonance with the applied laser beam. Scanning the resonance is typically performed by changing the potential which is applied to the charge exchange cell or the detection region.}
    \label{fig:beamline}
\end{figure}
\begin{figure}[ht]
    \centering
        \includegraphics[width=0.98\textwidth]{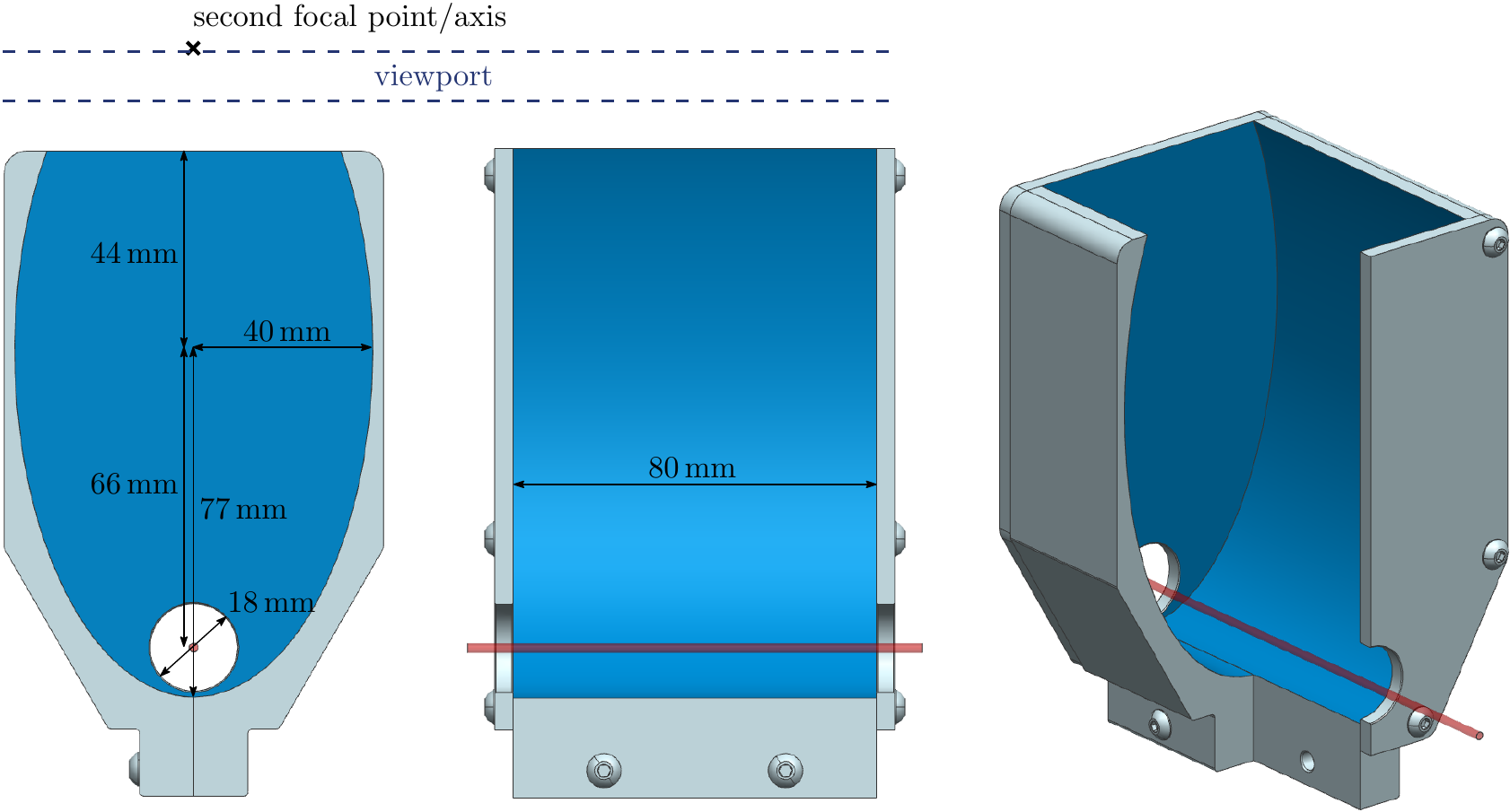}
    \caption{Left, center: The design parameters and dimensions of the oval mirror, visualized in two perpendicular cuts through the system. The surfaces colored in blue represent the mirrors, and the red line shows the beam axis. The position of the viewport and the second focal point just above is indicated. Right: A rendered computer-assisted drawing of the complete 4$\pi$ mirror with a cutout.}
    \label{fig:4picad}
\end{figure}
Laser spectroscopy is a renowned tool to investigate the electronic structure of ions and atoms, and to extract nuclear ground-state properties from optical spectra. By resolving the hyperfine structure and the isotope shift of an atomic transition, the nuclear spin and moments of the nucleus as well as its charge radius can be extracted \cite{Otten.1989}. This becomes increasingly interesting for rare isotopes with short lifetimes and low production rates at online facilities. Collinear laser spectroscopy is one of the working horses in this field \cite{Cheal.2010,Campbell.2016}. 

The basic principle of collinear laser spectroscopy is to superimpose a fast (keV) ionic or atomic beam \cite{Wing.1976,Anton.1978} with a narrow-bandwith laser beam, which is tuned to match the frequency of an optical transition. Either the laser frequency is scanned or the beam velocity is modified with a variable voltage to identify the resonance condition. In the latter case, the Doppler shift changes, leading to the same result as a frequency scan. This approach --called Doppler-tuning-- offers several practical advantages for example in the operation of the laser at a fixed frequency and in faster scanning procedures, and is therefore commonly applied. If the laser frequency matches the Doppler-shifted transition frequency, the atoms or ions under investigation are excited and spontaneously emit fluorescence photons. These photons are collected by a single photon detector, which are situated outside the vacuum chamber \cite{Neugart.1981}. Typically, photomultiplier tubes (PMTs) with bialkali photocathodes are used. A scheme of a collinear beam line using fluorescence detection is shown in Fig.\,\ref{fig:beamline}. 

Especially for experiments on species with low production rates, the detection efficiency is crucial. Furthermore, the significance of a resonance signal in a photon counting experiment is subject to the rate of background photons which inevitably reach the detector also under non-resonant conditions. Current state-of-the art systems are usually based on lense assemblies \cite{Kreim.2014} which are versatile in their application but have rather small solid angles of detection. Previously, also combined mirror-lens systems in connection with light guides were used but only applicable in the infrared regime \cite{Mueller.1983}. Here, we present a recently developed fluorescence-photon detection system, shown in Fig.\,\ref{fig:beamline}. Instead of lenses, an oval-shaped mirror system covers a portion of the beam and thus an almost 4$\pi$ photon-detection solid angle for the inside-vacuum part can be reached. The collected photons exit the beam line through a viewport and need to be guided into a photomultiplier tube. We analyze different solutions for such guides, focussing on compound parabolic concentrators which make use of passive non-linear mirror elements for background-light suppression. 

The first prototype was implemented at TRIGA-LASER \cite{Ketelaer.2008} for offline measurements on stable calcium isotopes \cite{Gorges.2015}. A revised version is already in use at the COALA-experiment in Darmstadt \cite{Konig.2020} and served well for laser-induced high-voltage measurements with calcium  \cite{Kramer.2018, Mueller.2020}, indium \cite{Konig.2020}, and isotope shift measurements on barium \cite{Imgram.2019}. Two more systems were built. One of them is already in use at the BECOLA facility \cite{Minamisono.2013} at the National Superconducting Laboratory at Michigan State University. Here, with the system implemented, it was possible to observe the D2 transition in $^{36}$Ca with a rate of low as 20-30 ions per second \cite{Miller.2019}. Another system is being deployed for laser spectroscopic measurements at Argonne National Laboratory, investigating the light proton-halo candidate $^8$B and palladium isotopes in the near future \cite{Maass.2017}.

\section{The 4$\pi$ oval mirror} \label{chapter:oval_mirr}
\begin{figure}[ht]
    \centering
        \includegraphics[width=0.35\textwidth]{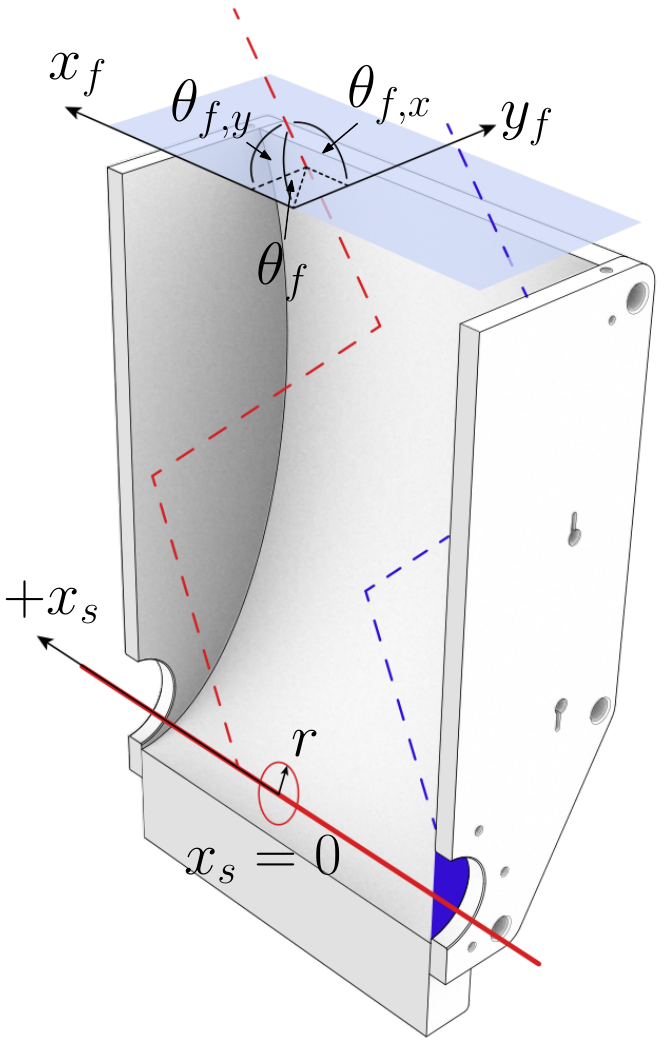}
    \caption{A display of two light sources which are used in the simulations of the 4$\pi$-mirror:  Signal photons (red) emerge from the beam with a given radius $r$, while stray light photons (blue) start from a disc in the entrance (or exit) aperture of the 4$\pi$ mirror, with a radius of $9\,$mm. Each photon's starting position $x_s$ and $r$ is recorded together with its end point ($x_f$,$y_f$) and angle $\theta_f$ on the viewport surface.}
    \label{fig:sim_coord}
\end{figure}
As soon as laser and particle beam are overlapped and under resonance conditions, photons are emitted from the corresponding beam volume. Ideally, both laser and particle beam have a Gaussian-shaped cross section, which results in a Gaussian-shaped overlap cross section as well, and both beams have similar widths (or diameters). The light source for signal photons in collinear laser spectroscopy can then be understood as the intersection volume of laser and particle beam. The adjustment between the two beams is critical not only to guarantee a long overlap distance, but also to control the Doppler-shift.

In principle, the emission is anisotropic and polarization-dependent, which can lead to small shifts of the resonance center especially when detecting only photons from a small solid angle. For the system which is introduced here, however, the large detection solid angle will counteract such effects and for simulations and experiments with typical resolution it is adequate to assume an isotropic photon emission. This has been demonstrated in Ref.\,\cite{Mueller.2020}. 

Naturally, also unwanted background photons exist, which do not result from the investigated excitation process. Two types of background photons have to be distinguished. Beam-related background emerges from the beam itself, and is the result of de-excitation processes which are not driven resonantly by the examination laser. It is commonly present after charge-exchange processes, when recombined atoms decay to their neutral ground state or after collisional excitation on residual gas atoms. This background is spatially indistinguishable from signal photons, but in some specific cases can be reduced with color filters or beam purification methods.

Stray light is laser light which is diffracted at apertures along the path. It can be mitigated with a well aligned optical setup including sets of diaphragms along the path. Since it does not emerge directly from the beam, its spatial and angular distribution differs from signal light. In simulations, we investigate these differences and how to use them to passively suppress stray light while forwarding signal photons to the PMTs. These single-photon counting PMTs are situated behind large-area view ports which have high optical transmission coefficients. The light guiding systems from the viewport surface to the PMTs will be discussed in the next chapter.

First, we show the mechanical layout of the in-vacuum mirror setup. Based on the derived dimensions, we present a series of simulations which visualize the in-vacuum photon transport from the beam to the viewport in our setup. 

\subsection{Layout of the mirror}
The first step of this photon guiding system is the oval mirror which is situated inside the vacuum beam line. As shown in Fig.~\ref{fig:4picad}, it consists of a mirror with oval cross section, which is extruded 80\,mm along the beam axis. One focal axis is coincident with the beam axis, the second axis lies just above the large-area view port window of the vacuum chamber. The semi-axes of the ellipse are 40\,mm and 77\,mm, and it is cut on the view-port-facing side, 110\,mm above the focal point which coincides with the beam axis. The faces of the extrusion are capped with flat mirrors, which have openings where the particle and laser beams enter and leave the oval mirror system. All parts were milled from plain aluminum and installed in a custom-made vacuum chamber, based on a standard conflat-160 cross with a shorter (112.5\,mm) extrusion which is equipped with a conflat-160 $d=100$\,mm view port. The size and geometry of the in-vacuum assembly is constrained by tube diameters and the distance between axis and view port. Simulations showed that, within reasonable limits, we do not expect fundamental performance differences for differently scaled systems.

The size of the oval mirror is constrained by the diameter of the vacuum tube and the disproportional increase in effort and cost to use larger or rectangular windows. Ultimately, also the light collection on the air-side is not trivial and becomes more difficult with larger areas. The assembly provides a 4$\pi$ collection of photons that are emitted from the beam axis to outside vacuum, neglecting the beam entry and exit openings. In the next subsections, the performance of the system is exemplified with the help of ray-tracing simulations.

\subsection{Detection solid angle}
\begin{figure}[ht]
    \centering
        \includegraphics[width=0.75\textwidth]{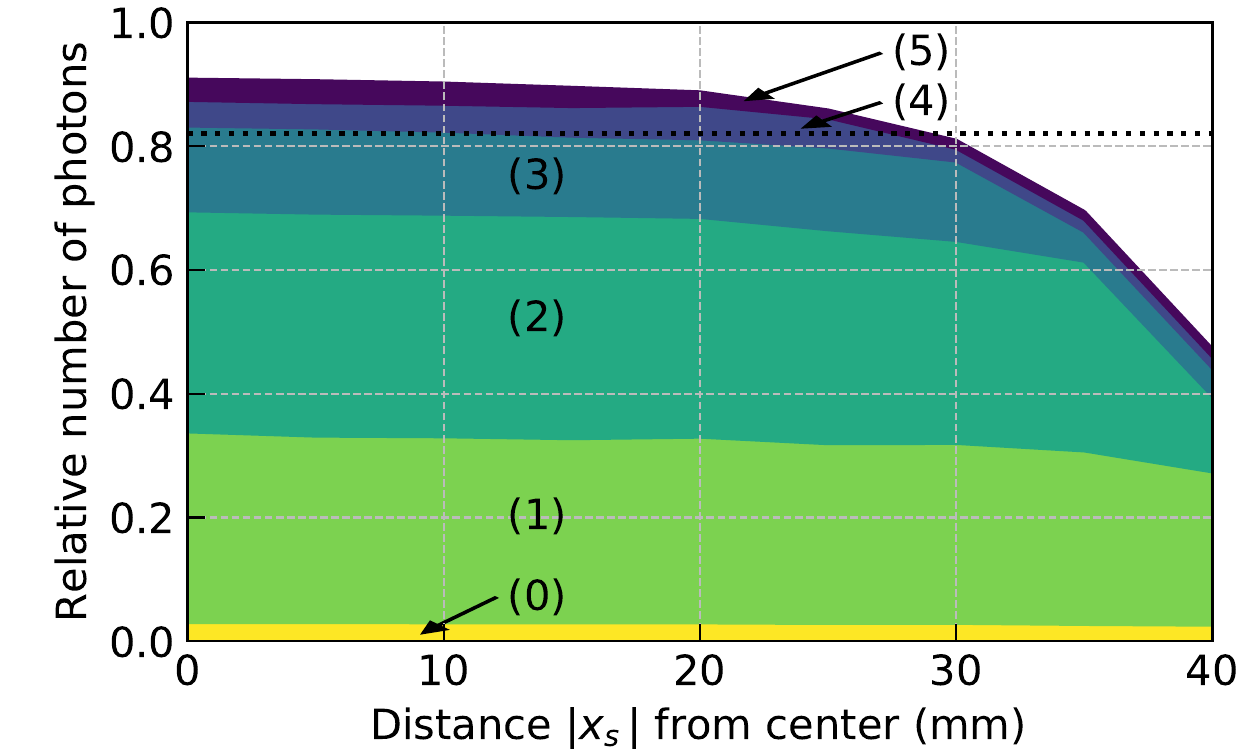}
    \caption{Photons emerging from a beam with $r=0$ are simulated, and the probability of them reaching the viewport surface is plotted with respect to their starting position $\left|x_s\right|$ on the beam axis. The reflectivity is set to $R=1$. The quota becomes smaller closer to the entrance (or exit) opening, but is 82\,\% in average. In color, the number of surface interactions of the photons is plotted, and the number is given in brackets. Most photons are reflected once or twice in the mirror, averaging to 1.8 reflections.}
    \label{fig:4pi}
\end{figure}
The primary purpose of the mirror is to cover the maximum solid angle for guiding photons that are sent out from the beam axis to the outside surface of the viewport. As shown in red in Fig.~\ref{fig:sim_coord}, the light source is modelled as a cylinder with radius $r$ which coincides with the axis of the ellipse and reaches from the entry at $x=-40$\,mm to the exit opening at $x=40$\,mm. In the simulations performed here, it is sufficient to analyze the dependency of the absolute value $\left| x \right|$ due to the symmetry of the mirror. Figure\,\ref{fig:4pi} shows the relative amount of photons which are transmitted through the viewport (which is set to zero absorption here). The dependence on $\left| x \right|$ can be understood since close to the entrance and exit holes, the probability of losing the photon becomes larger. Integrated over the whole beam length of 80\,mm, a total of 82\,\% of all photons are transmitted.

Moreover, in this simulation, also the number of surface intersections is counted. The reflectivity of the mirrors in the oval mirror is set to $R=1$, so no loss is calculated. However in reality $R<1$ must be considered, and the number of reflections for each photon is critical to evaluate the need for high-reflectivity mirrors. Most photons are reflected once (29\,\%) or twice (33\,\%) inside the oval mirror, averaging to 1.8 reflections. A typically reflection pattern includes one reflection on a flat and one on a curved mirror, like the one shown in Fig.~\ref{fig:sim_coord}.

\subsection{Output pattern}
\begin{figure}[ht]
    \centering
        \includegraphics[width=0.75\textwidth]{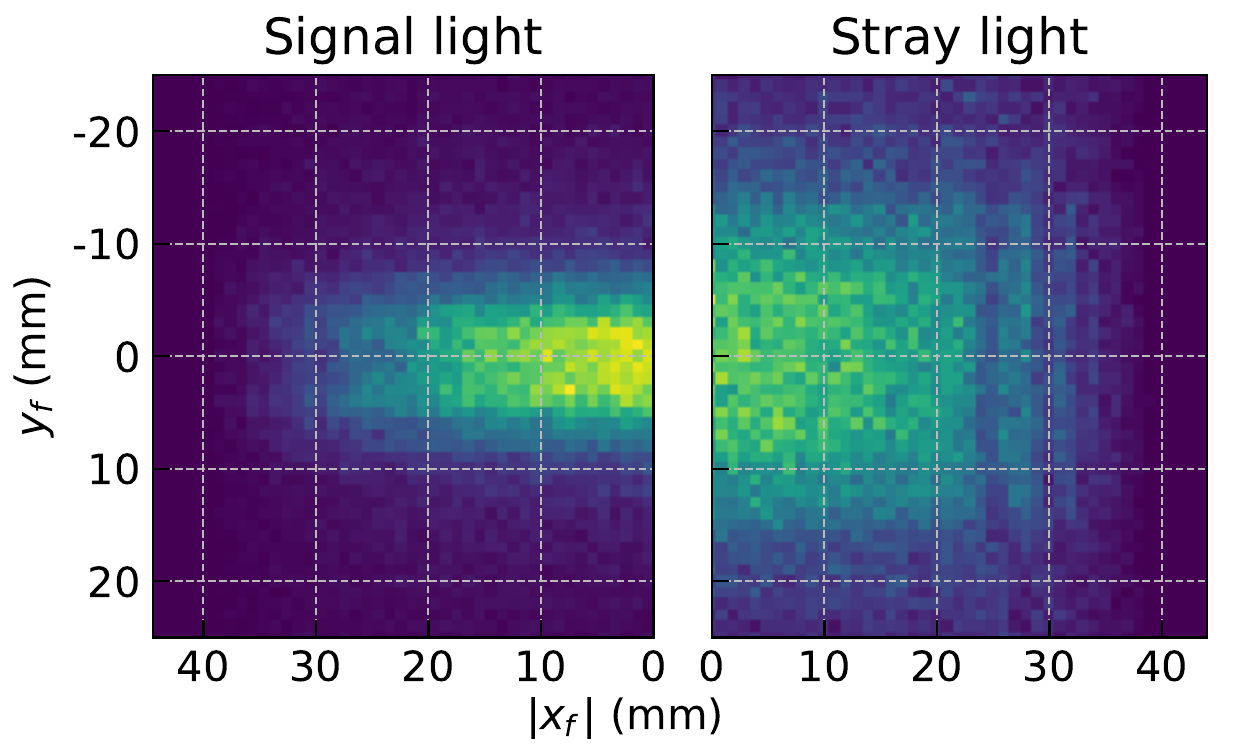}
    \caption{The spatial output pattern in the viewport plane of a simulation with a beam diameter of $r=1$\,mm and $r=9$\,mm background discs is plotted. The spatial distribution of signal light is shown in the left subplot, the stray light in the right. Since the simulation setup, and thus the output, is symmetric, both $x$-axes give the distance from the center of the 4$\pi$-mirror. Please refer to Fig.\,\ref{fig:sim_coord} for a visualization of the simulation light sources, including the parameters used.}
    \label{fig:acceptance_spat}
\end{figure}
\begin{figure}[ht]
    \centering
        \includegraphics[width=0.75\textwidth]{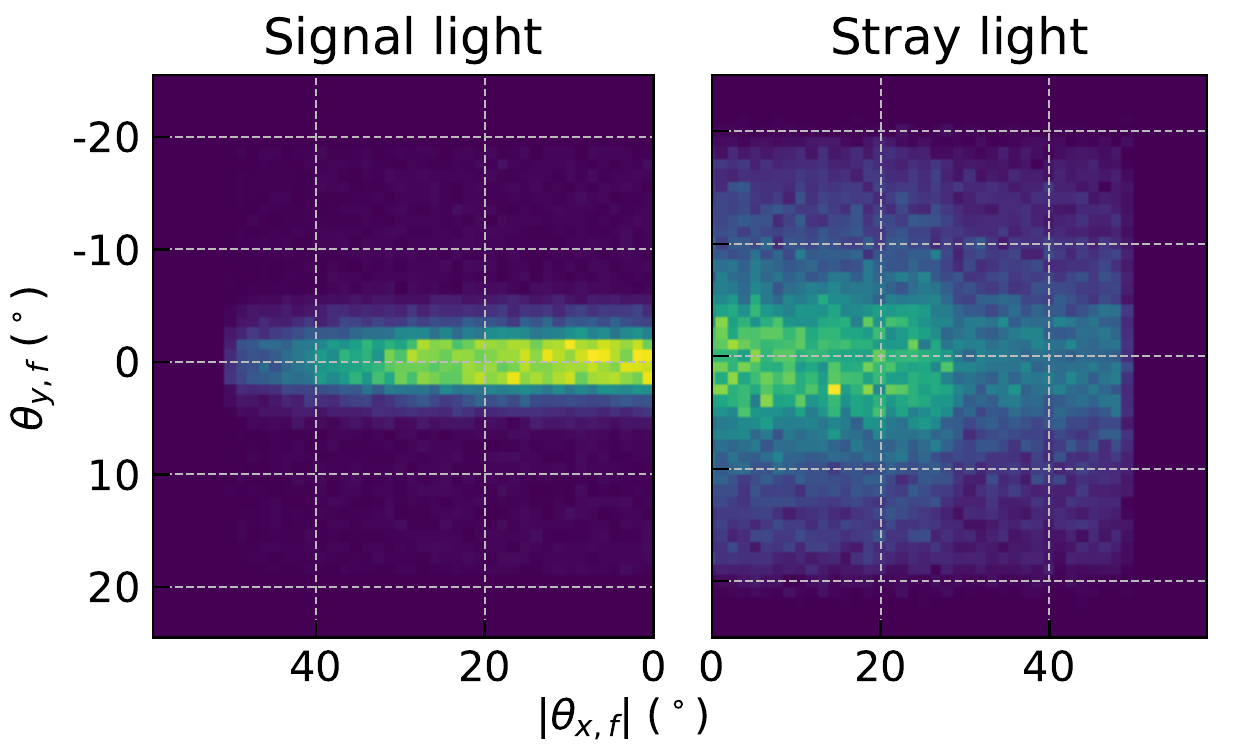}
    \caption{The angular output pattern in the viewport plane of a simulation with a beam diameter of $r=1$\,mm and $r=9$\,mm background discs is plotted. The angular distribution of signal light is shown in the left subplot, the stray light in the right. Since the simulation setup, and thus the output, is symmetric, both $\theta_{x,f}$-axes give the absolute angle in $x$-direction. Please refer to Fig.\,\ref{fig:sim_coord} for a visualization of the simulation, including the parameters used. Both plots use measurements in degrees, where $\theta_{y,f} = \theta_{x,f} = 0$ describes a beam which exits the viewport perpendicular.}
    \label{fig:acceptance_ang}
\end{figure}

\begin{figure}[ht]
    \centering
        \includegraphics[width=1\textwidth]{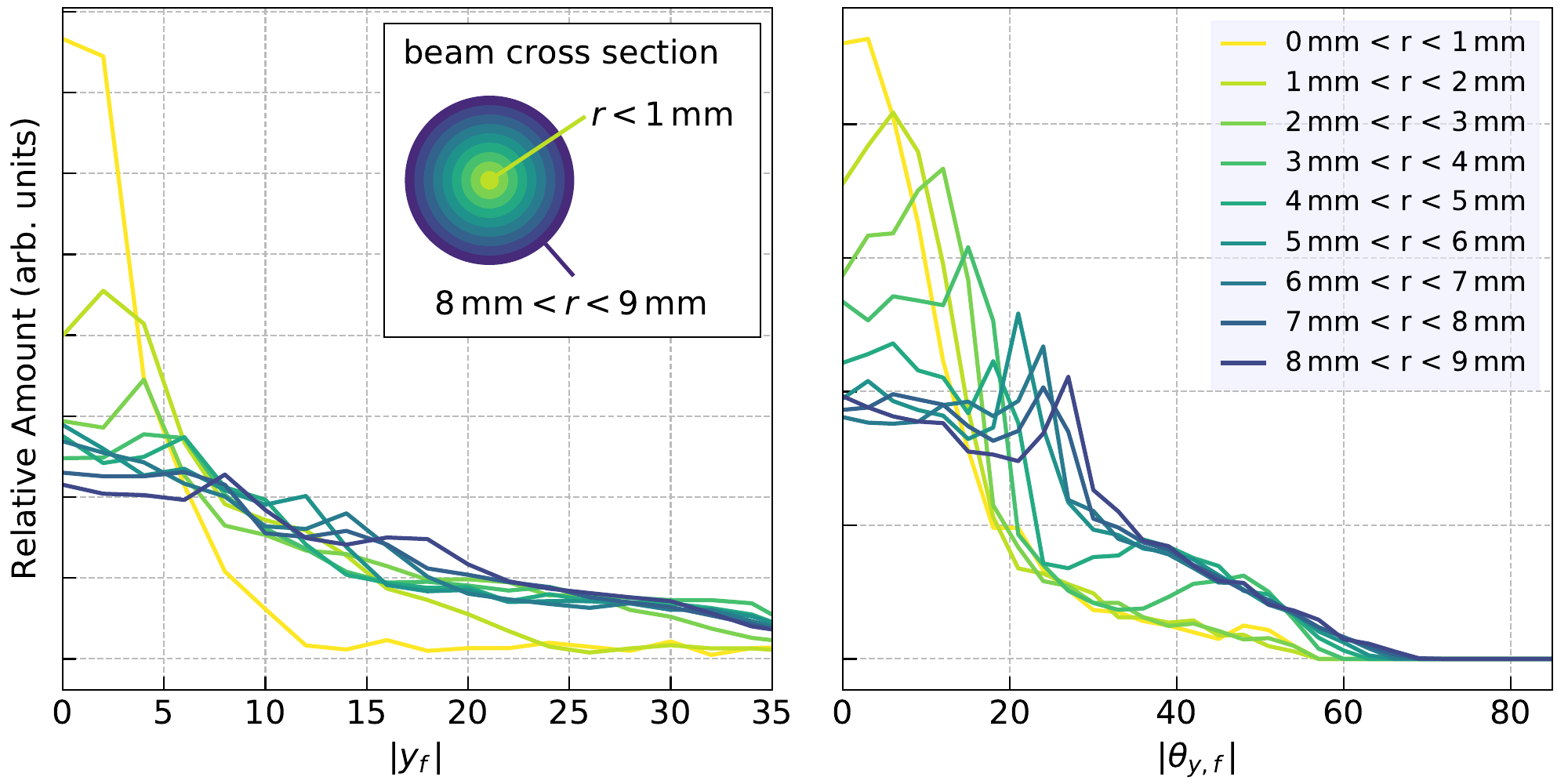}
    \caption{The spatial (left) and angular spread (right) of signal photons in the viewport plane with respect to the $y$-axis for different starting beam diameters $r$. This visualizes a vertical cross section through the histograms displayed in Fig.\,\ref{fig:acceptance_spat} and \ref{fig:acceptance_ang} with varying radius $r$. Due to the symmetry of the 4$\pi$-mirror, the absolute value is used. For the angle, the spread increases slowly, allowing a discrimination between photons generated at beams with two different diameters. In the spatial plain, the difference vanishes quickly above beam diameters >6\,mm, making discrimination based on spatial properties challenging. See text for details.}
    \label{fig:diam}
\end{figure}
The raytracing simulation allows a detailed investigation of the light collection system by examining the spatial and angular properties of the transmitted photons in the viewport plane. The obtained information will be used in the next chapter to choose an appropriate lightguide that transmits the photons to the PMTs. This is especially important when modelling both signal light and stray light to identify specific differences that can be used for passive discrimination. Signal light, by definition, originates from the beam which passes along the axis in random angles. Stray light is more difficult to model, since it is a result of diffraction processes that happen on the beam path mostly before the mirror system. As a least-assumptive model, we chose beam starting positions which are distributed in a disc shape, which is positioned inside the entry opening with the full diameter of 18\,mm. The photons are emitted isotropically into the 2\,$\pi$ angular range directing towards the mirror system. The background light source is displayed in blue color in Fig.\,\ref{fig:sim_coord} as well. The signal light source has a realistic radius of $r=1$\,mm in the simulations. However, the ratio of photons in the stray light and the beam light can never be assumed correctly, since it strongly depends on the experimental conditions.

The results of the simulation are shown in Fig.~\ref{fig:acceptance_spat} and Fig.~\ref{fig:acceptance_ang} for the spatial distribution and the angular distribution of photons on the view port plane, respectively. Both plots are cut symmetrically into two parts at $x_f = \theta_{x,f} = 0$ and show the output pattern for signal light and stray light, allowing an easy comparison.
The relative intensity is color-coded in the 2D histogram, giving a perception of where and at which angle the photons arrive at the view-port plane. The stray light exhibits a broader distribution both in the angular and in the spatial plane. Thus, it reaches the viewport in larger angles and more scattered across the surface. 

We realized that the spatial width of the photon distribution shown in the right half strongly depends on the diameter of the disc-shaped background light source and less on its placement. Thus, we concluded that the essential difference between the stray light and signal light distribution is connected with the difference in axial diameter of the two sources. To visualize this, a simulation was performed with an $r=9$\,mm beam, and the axial distance $r$ and the output parameters of each photon in the viewport plane is recorded. Then, ring-shaped segments $r_1 \leq r \leq r_2$ are extracted and the output spatial and angular distribution is plotted for each segment individually in Fig.~\ref{fig:diam}. To simplify the display, only the final $y-$coordinate and the respective angle is plotted (see Fig.~\ref{fig:sim_coord}). The plots indicate that for small beam diameters, a large offset in the viewport plane is unlikely but it increases steeply before plateauing for 6\,mm diameter beams. Thus, it seems possible to implement spatial background suppression, e.g. with a slit system, but only when the beam is sufficiently small in diameter. 

For the exit angle distribution, the spread is more gradual towards higher beam diameter, making a discrimination based on incident angle realizable as long as the stray light is generated in volume with larger diameter than the beam. Based on these findings, in the following chapter, we propose a solution to forward a large portion of the photons exiting the view port to the photomultiplier tubes, while discriminating stray light photons based on their angular distribution.

\section{Lightguides} \label{chapter:lightguides}
\begin{figure}[ht]
    \centering
        \includegraphics[width=0.7\textwidth]{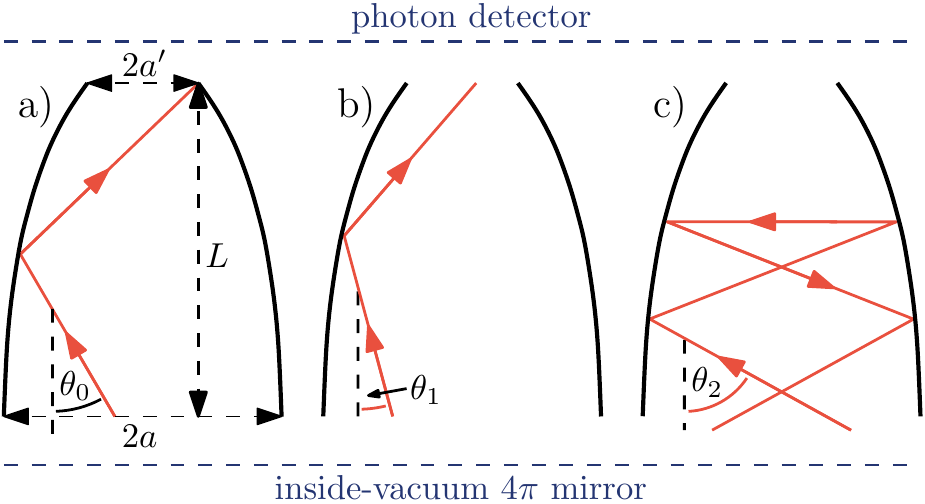}
    \caption{A compound parabolic concentrator (CPC) is defined by its cutoff angle $\theta_0$ and the size of the input and output apertures $2a$ and $2a'$, as denoted in (a). A beam with a smaller angle $\theta_1<\theta$ is  forwarded to the receiver, as shown for $\theta_1<\theta$ in (b).  Beams with larger angles $\theta_2>\theta$ are back-reflected multiple times and leave the CPC (c).}
    \label{fig:cpcs}
\end{figure}
Photons are reflected from the oval mirror to the surface of the viewport. Here, they impinge with the large spatial and angular distribution, which were simulated and are depicted in Figs.~\ref{fig:acceptance_spat} and~\ref{fig:acceptance_ang}. However, standard photomultiplier tubes, which detect the photons, generally have a small active area that does not cover the size of the spatial photon distribution at the viewport. In our case, the PMT active area is circular with $2a'=22$\,mm diameter. Thus, a second optical stage is required outside vacuum to guide the photons from the viewport area to the photomultiplier. Typically, the necessary conical reduction in cross-section that forwards photons by reflection results in a trade-off between spatial and angular acceptance.

A common tool to achieve this light concentration is a compound parabolic concentrator (CPC). It has the capability of reflecting incident radiation within wide limits to the absorber or detector. Therefore, their advantages are not limited to scientific detectors but they can also be used, e.g. as collectors of solar energy as pointed out in \cite{Winston.1974}. The function of a two-dimensional CPC is depicted in Fig.~\ref{fig:cpcs}. The parabolic cross section forwards angles which arrive at steep angles $\theta$ from the input to the output, but rejects light with shallow incident angles. This is ideal for our system, where we concluded that unwanted background light has a tendency to follow more shallow ray paths through the viewport window, as shown before in Figs.~\ref{fig:acceptance_spat} and~\ref{fig:acceptance_ang}. For a CPC with input aperture $2a$ and output aperture of $2a'$, its length $L$ is determined by
\begin{align} \label{eq:cpclength}
    L = \frac{a'(1+\sin{\theta_0})\cos{\theta_0}}{\sin^2{\theta_0}}
\end{align}
with the half angle $\theta_0$. This angle defines the angular cut-off and is correlated with the aperture dimensions via
\begin{align} \label{eq:cpcsize}
    \frac{a}{a'} = \frac{1}{\sin{\theta}}~.
\end{align}
Thus, the size of the receiving end aperture and the cutoff angle $\theta_0$ determines all other dimensions, such as the length or the entry aperture size. The construction pattern for a 2D-CPC can be found in literature~\cite{Chaves.2008}. 

A spherical circular CPC is generated by rotating the two-dimensional parabolic cross section around the symmetry axis, resulting in a homogeneously-shaped circular system. Such systems are the best approximation to a stereoscopic concentrator but they are difficult to fabricate since the reflective surfaces are bent along two axes, making any surface treatment challenging. Thus, we implemented a simpler, approximated shape that we named quadratic parabolic concentrator (QPC) which is shown in the figures of this article, e.g. Fig.\,\ref{fig:sim_coord_cpc}. It is generated by joining four quarters which have a parabolic cross section that are extruded linearly. 

A QPC20 and a QPC30 with the cutoff angles of 20$^\circ$ and 30$^\circ$, respectively, were built. Here, we first compare the performance of a quadratic and circular concentrator, and then briefly outline on what basis the two explicit cutoff angles were chosen.

\subsection{Concentrator shape}
\begin{figure}[ht]
    \centering
        \includegraphics[width=0.4\textwidth]{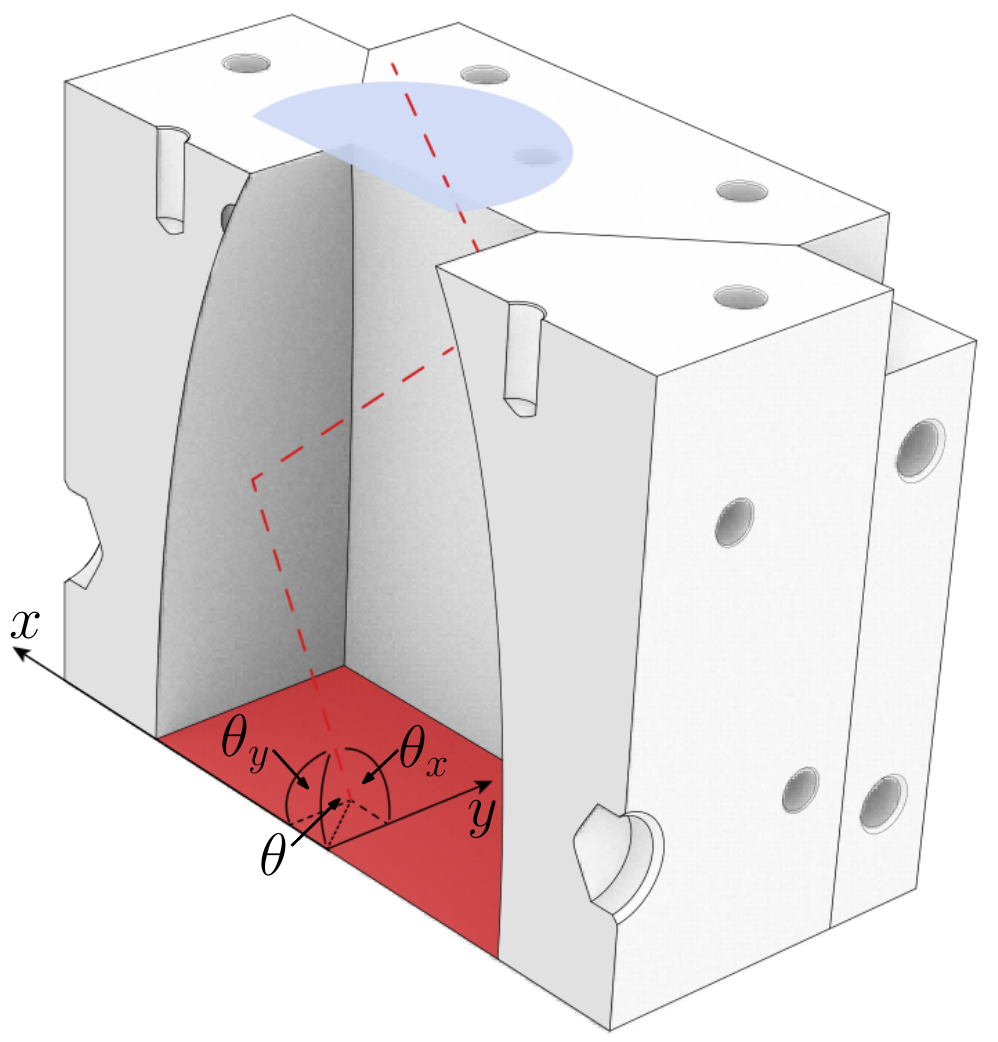}
    \caption{A display of the simulation parameters and coordinates used to characterize the different lightguides. The photons emerge from entrance area, and their starting position and angle is recorded together with the information if they reached the PMT surface or not.}
    \label{fig:sim_coord_cpc}
\end{figure}
\begin{figure}[ht]
    \centering
        \includegraphics[width=1\textwidth]{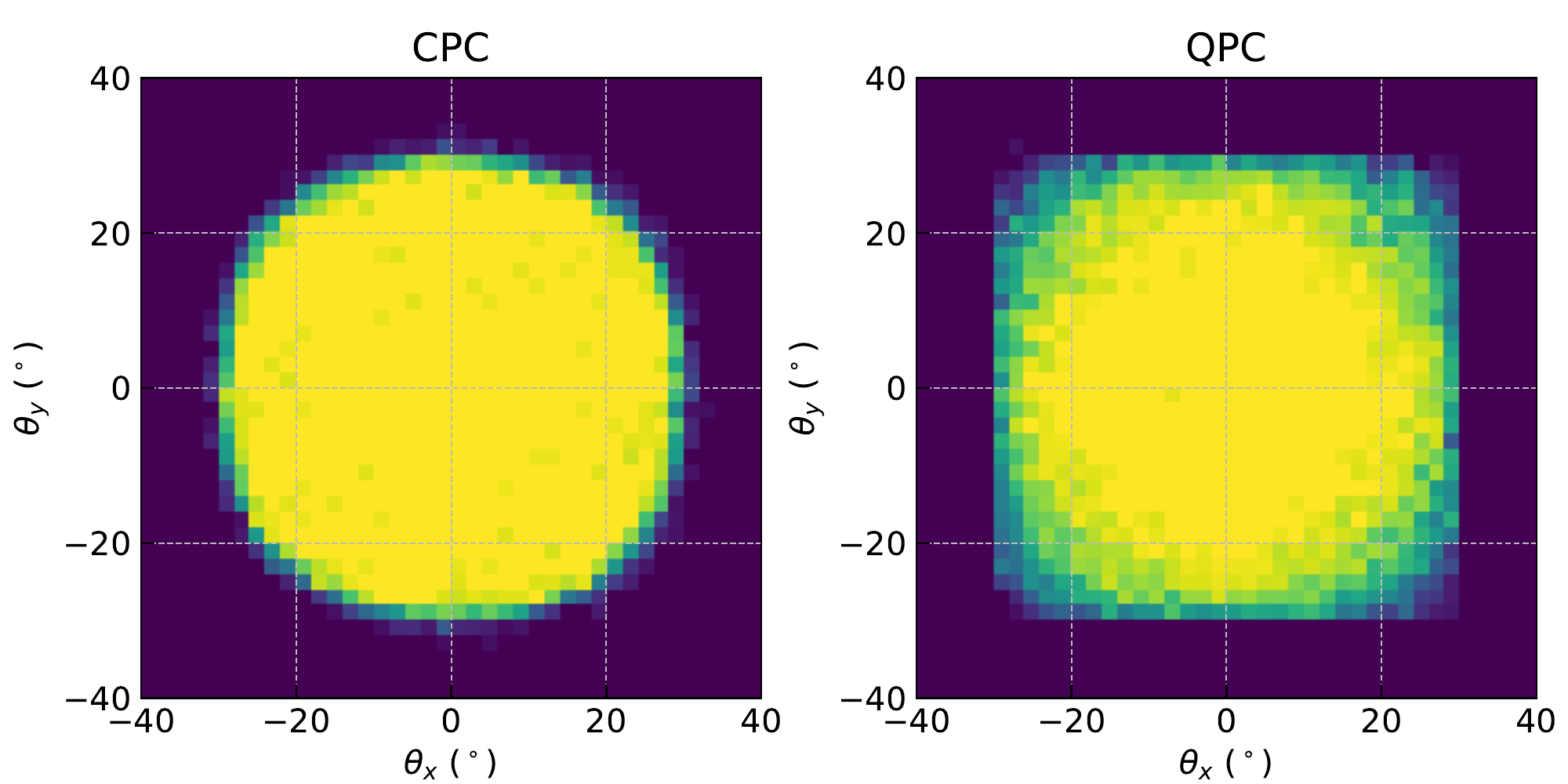}
    \caption{A circular (a) and a quadratic (b) parabolic concentrator with cutoff angle of $\theta_0 = 30^\circ$ are compared by investigating the transmission probability of light that is emitted in random angles from the viewport plane. Both models show the expected cutoff and a close to perfect transmission for beams with angles $\theta < \theta_0$.}
    \label{fig:cpc}
\end{figure}
The performance of a circular and a quadratic parabolic concentrator can be compared in raytracing simulations. Figure\,\ref{fig:sim_coord_cpc} gives an overview over the simulation in the case of a QPC30, but the given angles and dimensions are similar in both setups. Systems with $\theta_0 = 30^\circ$ are compared, but there is no significant difference when using other cutoff angles. The respective full input window area is populated with a light source that emits photons in random directions, and the photons which reach the PMT detector plane are counted. In Fig.~\ref{fig:cpc}, a normalized histogram shows the distribution of beam starting angles that reach the exit of the lightguide. The CPC presents a smooth, almost perfect efficiency for any photon within the acceptance angle, while the QPC has diffused edges which reflect the quadratic shape. However, the total number of transported photons is almost identical (~25\,\%), while the entrance window is smaller for the CPC due to its circular shape. The average number of surface reflections is 2.2 for both systems.

From this simulation one can conclude that QPCs are well-suited for the application in the fluorescence detection system since their performance is similar to CPCs, while being easier to implement due to surfaces which are only curved with respect to one axis.

\subsection{Choice of cutoff angle}
\begin{figure}[ht]
    \centering
        \includegraphics[width=1\textwidth]{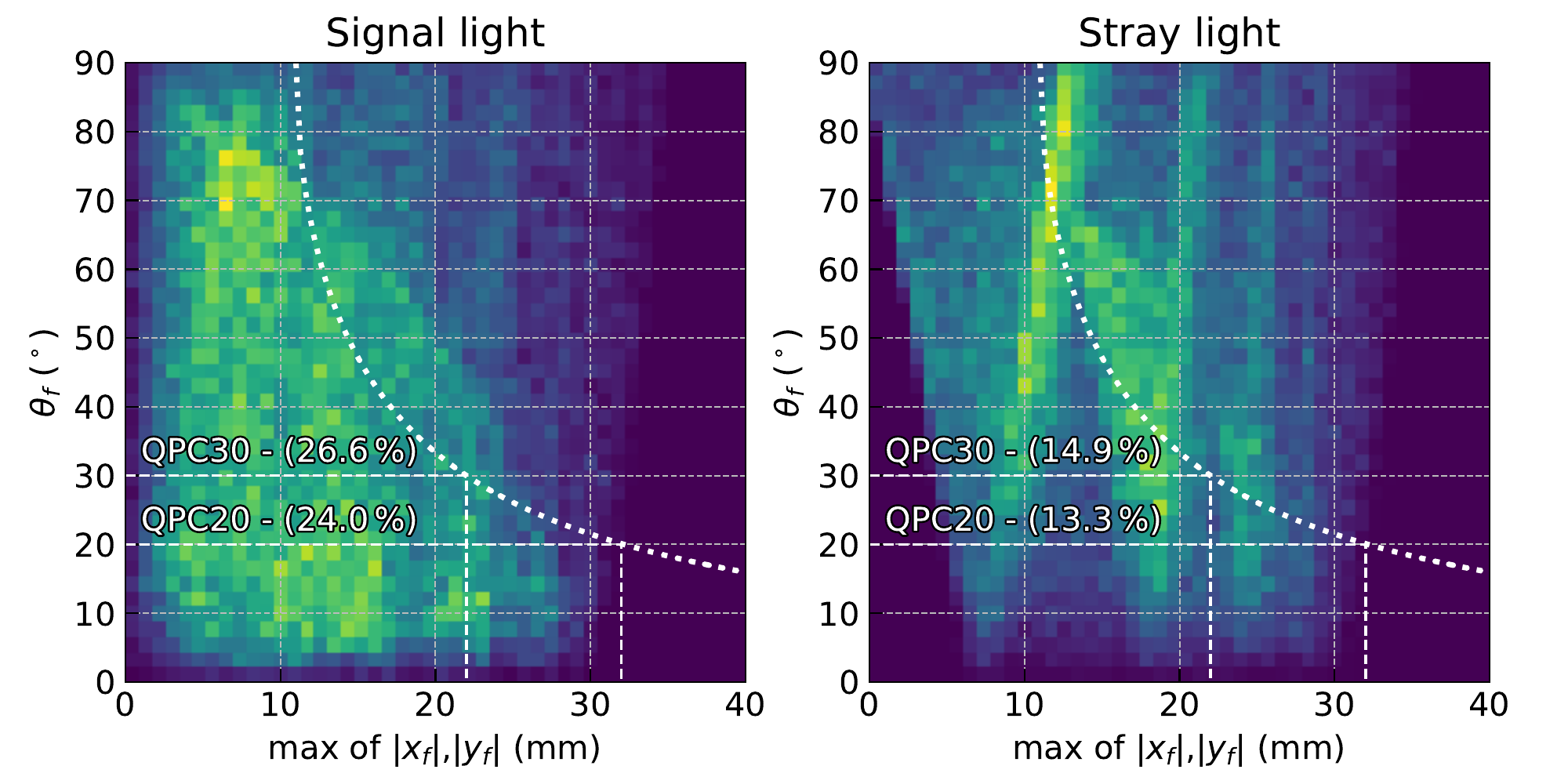}
    \caption{The two-dimensional histograms show the output of the 4$\pi$-mirror in the viewport plane for signal light from an $r=1\,$mm beam and stray light discs. Each photon's exit angle $\theta_f$ is recorded together with its position $\left( x_f,y_f \right)$. Since a QPC is centered on the window and quadratic, a photon will only enter if the larger absolute value of the two spatial coordinates is smaller than the QPC edge length $a$. The correlation between $a$ and $\theta_0$ given in Eq.~\ref{eq:cpcsize} is drawn as white dotted line, and the input area of a QPC30 and a QPC20 are denoted with the dashed rectangles. The percentages give acceptance of photons in the respective area.}
    \label{fig:choice}
\end{figure}
\begin{figure}[ht]
    \centering
        \includegraphics[width=.5\textwidth]{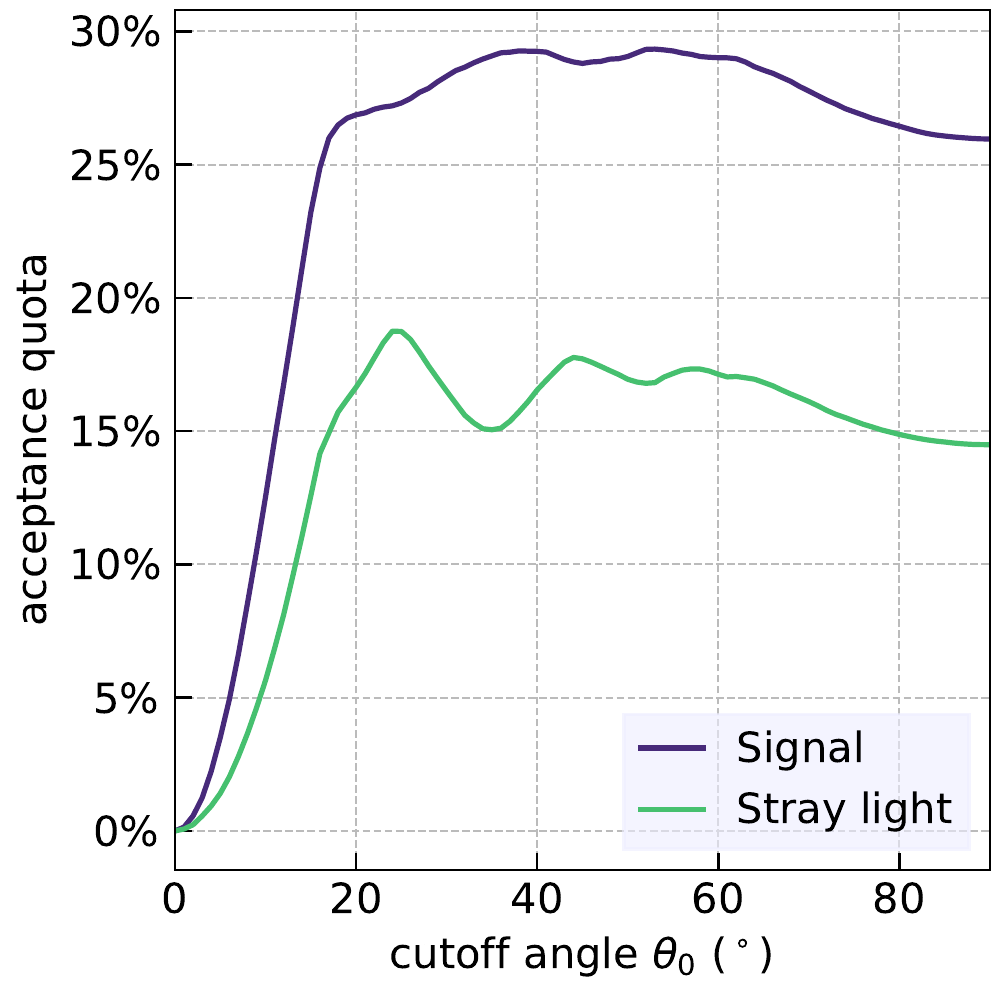}
    \caption{In a simplified display of the data presented in Fig.~\ref{fig:choice}, the quota of photons that are accepted by a QPC with a given cutoff angle $\theta_f$ is depicted. Since the acceptance area becomes smaller with less restrictive cutoff angles, the amount of photons which are collected stays more or less constant for $\theta > 20^\circ$.}
    \label{fig:numchoice}
\end{figure}
In the previous simulation, concentrators with $\theta_0 = 30^\circ$ cutoff angle were chosen as an example. Following Eq.~\ref{eq:cpcsize}, the choice of this cutoff angle determines the geometry of the system, since the employed PMT is best used with $a'=11$\,mm. Larger cutoff angles will forward a higher quota of photons to the PMT, but are linked with a smaller entrance area. To determine which cutoff angle is the optimum, the realistic output of the oval mirror needs to be taken into account. For this, Fig.~\ref{fig:choice} shows a visualization of the result of the simulation performed in Sec.~\ref{chapter:oval_mirr} from an $r=1$\,mm beam and 18\,mm disc-shaped stray light sources. Here, the higher absolute value of the $x_f$ and $y_f$ coordinate at the viewport plane is plotted versus the combined output angle $\theta_f$. Thus, the histogram gives the correlation between the spatial and the angular distribution of the photons exiting the 4$\pi$-mirror, which is significantly different for signal and stray light. To obtain the best ratio between signal and stray light photons, the area below 30$^\circ$  and 30\,mm offset seems most promising in terms of signal acceptance and stray light suppression. The two QPCs with 20$^\circ$ and 30$^\circ$ acceptance are plotted, as well as the curve which determines the ratio between angular cutoff and size which the QPCs follow per definition. The relative number of photons that lie within the respective QPC acceptance range is given as well for the two examples.

In principle, a QPC can have any cutoff angle from $\theta_f = 0^\circ$ to $90^\circ$, and with a given receiver size $a'$ the length $L$ changes according to Eq.~\ref{eq:cpclength}. Thus, we can  count the relative number of photons within the acceptance square in Fig.~\ref{fig:choice} for different cutoff angles to create Fig.~\ref{fig:numchoice}. Clearly, the QPC should be chosen with acceptance angle greater than 20$^\circ$ to avoid the area where the signal and background light still increases. Angles between 20$^\circ$ and 40$^\circ$ seem to be well suited; Larger cutoff angles lead to significantly reduced acceptance areas (which cut more than half of the 80\,mm oval mirror length). The extreme case is a 90$^\circ$ concentrator which by definition equals the PMT mounted directly to the viewport, since its length is 0 according to Eq.\ref{eq:cpclength}. Such a setup is easily realizible and its performance, according to Fig.~\ref{fig:numchoice} is not significantly worse than with real concentrators employed. However one should note that in reality, also the PMT has a limited angular acceptance. For the fluorescence detection system described here, we chose and fabricated QPCs with $\theta_f = 20^\circ$ and $30^\circ$ and also implemented a method to directly mount the PMT to the viewport. To differentiate between the different lightguide systems in the following, we name them QPC20, QPC30 and DC (direct collection).

\section{Composite system}
After characterizing the oval mirror and choosing appropriate variants of the concentrators, we can investigate the performance of the combined system based on findings in the simulations. Since the simulations per definition represent a perfect system, they can give an upper limit to the expected performance. 
\subsection{Forwarding efficiency}
\begin{table}[ht]
\centering
\caption{The simulated forwarding efficiency $\varepsilon_\textrm{forw}$ for the three lightguide systems, using an $r=1$\,mm beam and disc-shaped stray light sources as depicted in Fig.~\ref{fig:sim_coord}. The numbers give the percentage of sent-out photons that reach the detector surface within the given boundaries. As expected, all three systems (QPC20, QPC30 and DC) perform similar, with around 20\,\% forwarded signal light to the PMT.}
\label{tab:total_eff}
\begin{tabularx}{0.95\textwidth}{lXlXD..{2.1}XD..{3.0}}
\tls
\toprule\toprule
                                && \multicolumn{1}{l}{\text{Boundaries}}              && \multicolumn{1}{l}{\text{Signal}}                && \multicolumn{1}{l}{\text{Stray}} \\ 
                                &&               && \multicolumn{1}{c}{\textit{\%}}                 && \multicolumn{1}{c}{\textit{\%}} \\ 
\midrule
start                           &&                                                                  && 100.0                    && 100.0                          \\ 
viewport plane                  &&                                                                  && 79.1                     && 87.2                           \\ 
\midrule
\multirow{3}{0pt}{QPC20}        && max of $(\left|x_f\right|,\left|y_f\right|) < a = 32$\,mm        && 60.1                     && 62.2                           \\
                                && $\theta_f < \theta_0 = 20^\circ$                                 && 24.3                     && 15.3                           \\
                                && \textit{both}                                                    && 19.0                     && 11.6                           \\                                
\midrule
\multirow{3}{0pt}{QPC30}        && max of $(\left|x_f\right|,\left|y_f\right|) < a = 22$\,mm        && 42.0                     && 37.9                           \\
                                && $\theta_f < \theta_0 = 30^\circ$                                 && 40.2                     && 31.4                           \\ 
                                && \textit{both}                                                    && 21.0                     && 13.0                           \\ 
\midrule
DC                               &&  $\sqrt{x_f^2 + y_f^2} < r_\textrm{PMT} = 11\,$mm           && 18.8                     && 10.4                         \\
\bottomrule\bottomrule
\end{tabularx}
\end{table}
The simulation results for the oval mirror can be combined with the data from the concentrators to get an estimate of the forwarding efficiency $\varepsilon_\textrm{forw}$ of the composite system, which is defined as the total number of sent-out photons that reach the PMT. An $r=1$\,mm beam radius and a disc-shaped  lightsource with 18\,mm diameter are used as signal and stray light, respectively. The incoming photon distribution in the analysis plane of the viewport is then cropped to match the acceptance space of the different light guide systems. 

In Tab.\,\ref{tab:total_eff}, the results of the composite calculation are summarized. All three systems that are simulated show a similar performance: Around 20\,\% of all photons originating from the $r=1$\,mm beam are forwarded to the PMT, while only 10\,\% of the stray light is reaching the detector. The large acceptance area of the QPC20 is compensated by its more restrictive angular cutoff. Contrary, the PMT mounted directly to the viewport surface has a limited detection area, but has no direct restriction in angular space.

From the simulations performed here, evidently no superior concentrator design can be highlighted, based on the fact that all systems have a similar forwarding rate for signal photons. This parameter can be credibly extracted from the simulation, since the cylindrical shape of the beam is known and can be modelled. For the stray light, however, the disc-shaped light sources are a rough approximation to the complex scattering and diffraction processes that happen along the laser beam path. It can be expected that stray light enters the mirror with shallower angles, boosting the performance of the passive stray light suppression. The experimental investigations, presented in chapter \ref{sec:coala} confirm this assumption and show that in many cases, the use of QPCs is expedient and allows to acquire a meaningful signal faster.

Generally, the performance will depend strongly on the beam diameter and alignment, as well as on the exact parameters of the stray light. Thus, in the following chapters, results from different experimental conditions will be presented, and the influence of the mirror reflectivity is investigated.

\subsection{Reflectivity}
\begin{figure}[ht]
    \centering
        \includegraphics[width=.5\textwidth]{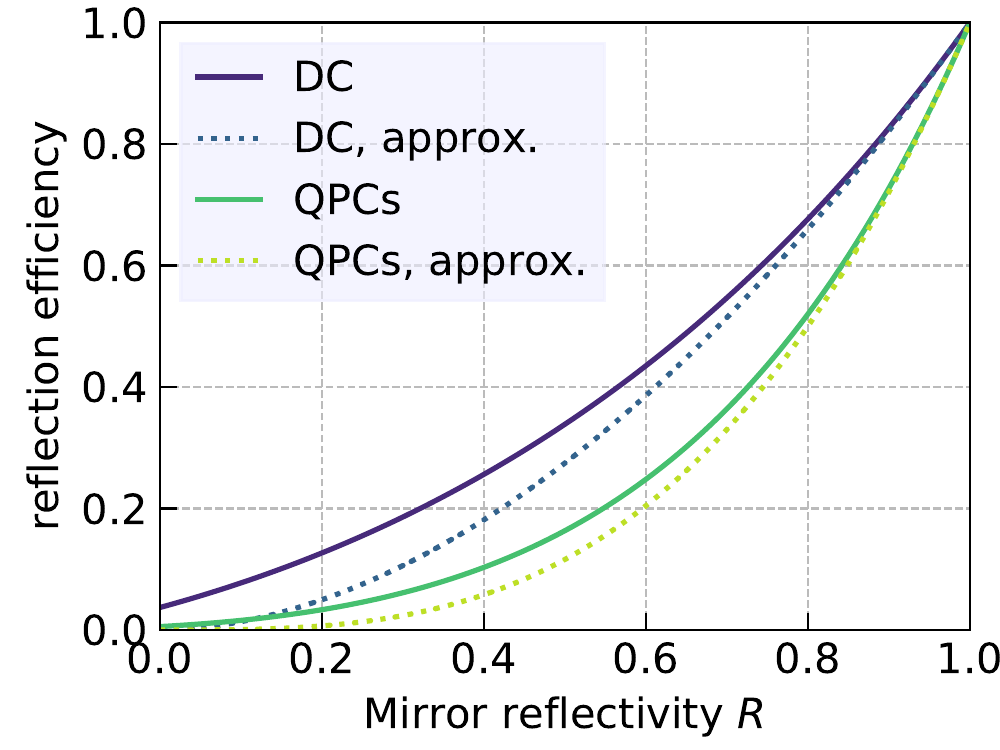}
    \caption{For given mirror reflectivity $R$, the reflection efficiency $\varepsilon_\mathrm{refl}$ gives the total number of photons which are forwarded through the system without absorption. The DC setup has a better $\varepsilon_\mathrm{refl}$ due to the average of 1.2 more reflections when using the QPCs. The exponential correlation emphasizes the need for highly-reflective mirror surfaces for the obal reflector and the QPC. The dotted curves give the simple exponential approximation based on the average reflection numbers for comparison.}
    \label{fig:refl_new}
\end{figure}
Since the fluorescence detection system is based on mirrors, their reflectivity for the wavelength of interest is a critical parameter for its performance. In the simulations performed so far, the number of reflections was recorded as well. In average, the signal photons are reflected 1.8 times in the oval mirror and 1.2 times in the QPCs while there is no additional reflection in the DC system. The attenuation can be approximated by
\begin{align}
    \varepsilon_\mathrm{refl} = R^{\bar{N}}
\end{align}
using the reflectivity $R$ of the surfaces and the average number of reflections $\bar{N}$. However, some photons reach the detector without a reflection and thus an exact approach is to introduce the reflectivity into the ray-tracing simulation. The resulting curves are displayed in Fig.~\ref{fig:refl_new}, including both the approximation and the ray-to-ray analysis. Since the reflection efficiency increases with higher mirror reflectivity, it is essential to have mirrors with high (>80\,\%) reflectivity, and even a small improvement can cause a relatively large net change. As a practical example, thin aluminium mirror sheets which are employed in both the QPCs and the oval mirror in one version of the fluorescence detection region have a reflectivity of $R=0.88$ at 400\,nm, which results in a reflection efficiency $\varepsilon_\textrm{refl} = 0.67$. In combination with the forwarding efficiency $\varepsilon_\textrm{forw} = 0.19$ of signal light in the QPC20 listed in Tab.~\ref{tab:total_eff}, a total efficiency $\epsilon_\textrm{tot} = 0.13$ is obtained. Effectively, every eighth photon which is sent out in the active area of the detection system reaches the PMT. Even with the unavoidable loss of photons in the multiple reflection process, a high efficiency can be achieved resulting in a high signal collection rate, which will be investigated on the basis of experiments later.

\subsection{Realization of mirror surfaces} \label{sec:mechanical}
\begin{figure}[ht]
    \centering
    \begin{subfigure}[c]{0.45\textwidth}
        \includegraphics[width=\textwidth]{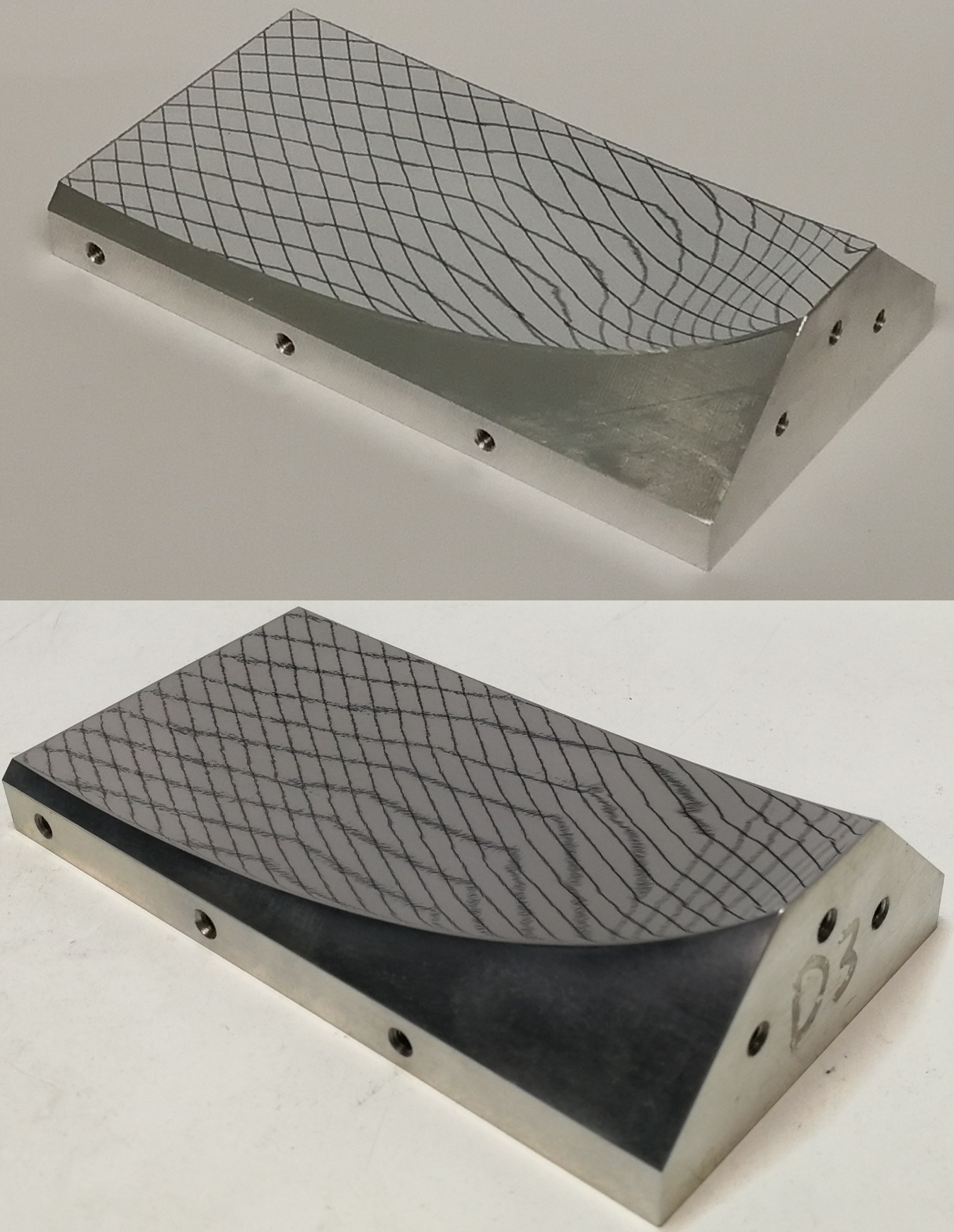}
        \caption{Segments of the CPC20. Top: Mirror sheet is glued to the surface. Bottom: Hand-polished segment. The mesh is a reflection which helps showing the mirror quality.}
        \label{fig:gull}
    \end{subfigure}
    \qquad
    \begin{subfigure}[c]{0.45\textwidth}
        \includegraphics[width=\textwidth]{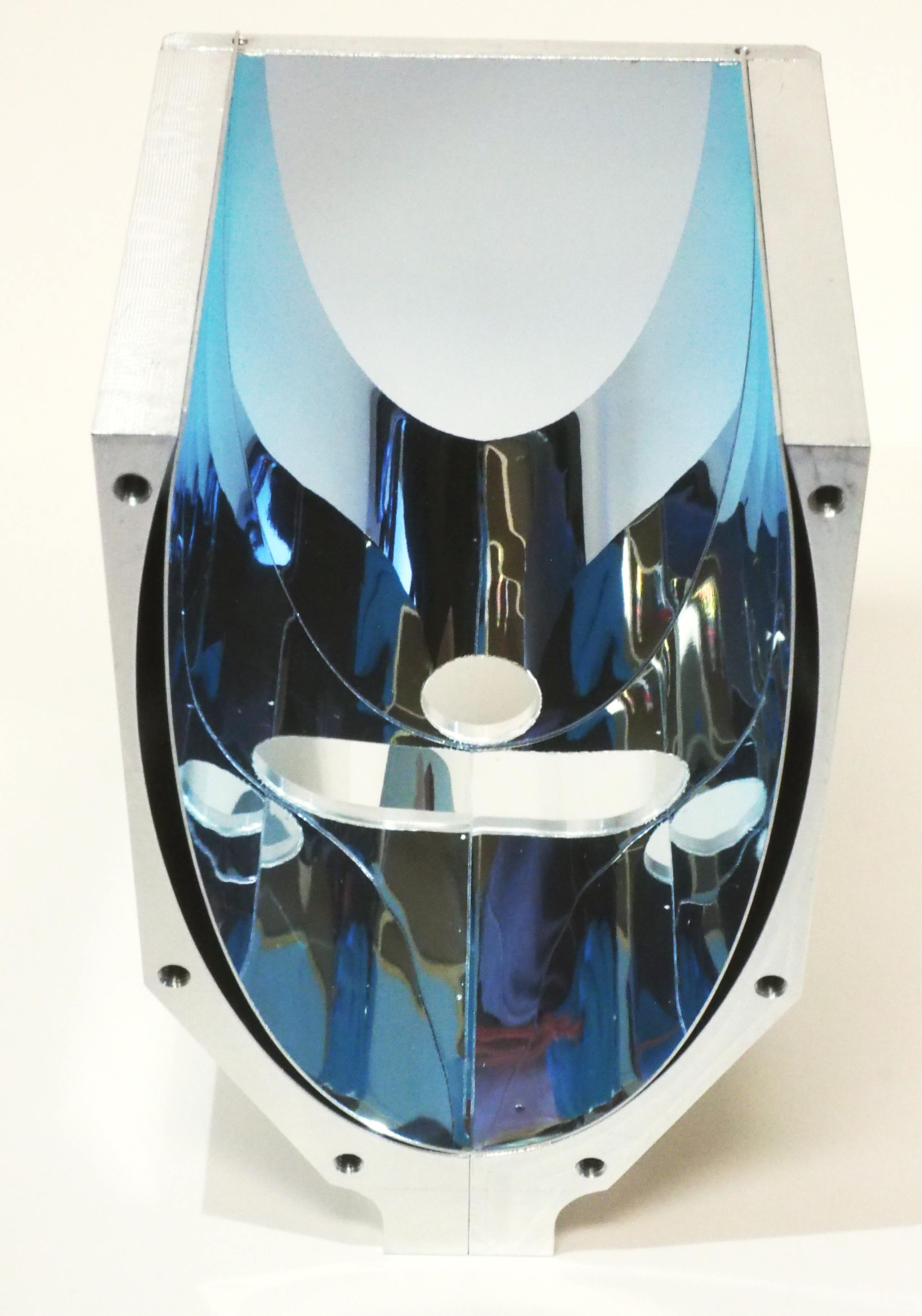}
        \caption{The oval mirror equipped with mirror sheets, which are not pressed into form yet. The front cap is missing here, while the back cap shows the aperture and its reflection.}
        \label{fig:tiger}
    \end{subfigure}
    \caption{Two different techniques were applied to create the mirror surfaces: Hand-polished mirrors, and sheet mirrors for industrial applications which are bent into shape. The hand-polished mirrors have a better deep-UV reflection at the cost of macroscopic optical defects, as visible in (a). The used sheet mirrors have excellent optical properties but they lack reflectivity in the UV. However, they can potentially be swapped with wavelength-specific mirrors.}\label{fig:mirrors}
\end{figure}
The realization of the large, bent mirror surfaces is particularly demanding, especially since the reflectivity of the mirrors is of utmost importance. At the same time, the in-vacuum mirror surfaces need to be electrical conductive to prevent charge-up effects when dealing with ionic beams. These two requirements, together with the special shape and large area, are not easy to fulfill especially when reaching out to far-UV (<250\,nm) regions. Generally, plain aluminum mirrors are a good choice since they have a high reflectivity of above 90\,\% in a broad spectral range reaching down to 190\,nm. For certain applications, higher performance could be achieved by utilizing wavelength-specific di-electric mirrors. Here, we focus on the implementation of two different mirror systems which are based on metallic aluminum.

In one approach, the curved mirror geometry was milled from plain aluminum pieces and subsequently hand-polished to optical quality. Figure~\ref{fig:mirrors}a shows a polished quarter of a QPC imaging a grid. Microscopic mirror quality is reached with reasonable effort, and reflectivity of $\sim 90\,\%$ is reached when testing with a laser beam at 400\,nm. However macroscopic errors (dents) from the milling and polishing process remain on the surface, which impairs the imaging properties of the mirror despite the high overall reflectivity.

This technique allows a perfect conductivity and vacuum compatibility, and the oval shape can be generated perfectly since polishing only effects the surface layer of the aluminum shape. However, bare aluminium corrodes when exposed to oxygen-containing air. This can be avoided for the inside mirrors by keeping them under vacuum or inert gas atmospheres. The QPCs, however, need to be coated with a thin quartz film that prevents corrosion while allowing UV reflection. This is feasible since conductivity or vacuum compatibility is not crucial for the QPCs since they are operated outside the vacuum system. However, this introduces some absorption when using low wavelengths. 

A second approach used thin (0.05\,mm) aluminum reflector profiles for industrial applications as mirrors. They proved to have an electrical conductivity and vacuum compatibility which is adequate for ion beam applications. Cut into the right 2D shape, they can be fitted and pressed (inside vacuum) or glued (outside vacuum) into the desired shape. Figure \ref{fig:mirrors}b shows a picture of one oval mirror which is equipped with such aluminium reflectors but not yet pressed into the oval cylinder. Compared to the hand-polished mirrors in Fig.\,\ref{fig:mirrors}a, the reflector profiles have excellent microscopic and macroscopic mirror quality. However, due to the coating, they are limited to wavelengths in the near UV regime. While the reflectivity conforms with the hand-polished mirrors at 400\,nm wavelength ($\sim 90\,\%$), at 250\,nm it has already dropped to 65\,\%. 

Since the inside vacuum mirror surfaces are simply pressed into place, they can be swapped easily to mirrors with reflectivity in the desired region. For example, spectroscopy in the infrared region could be optimized by using bare copper mirrors. On the UV side of the spectra, efforts are ongoing to enhance conductive coatings towards smaller wavelengths $\lambda < 250$\,nm.

\section{Experiment} \label{chapter:exp_res}
In collinear laser spectroscopy, fluorescence photons are counted while scanning the frequency of the laser that interacts with the particles in their moving frame of reference. When the laser is in resonance with an atomic or ionic level, the particles emit photons with a rate $\dot{R}$. Under resonance conditions, this rate depends on the power density of the laser field and the particle beam current. In experiments with rare, radioactive beams the current is usually low and limited by the production and transport to the experiment. The laser power can be scaled, but high powers lead to saturation and broadening effects in the resonance that interfere with the high precision that is usually needed. In the following, some fundamental parameters which are relevant to compare and evaluate experimental data are introduced before results are presented.

\subsection{Line shapes and strength normalization}
\begin{figure}[ht]
    \centering
        \includegraphics[width=.65\textwidth]{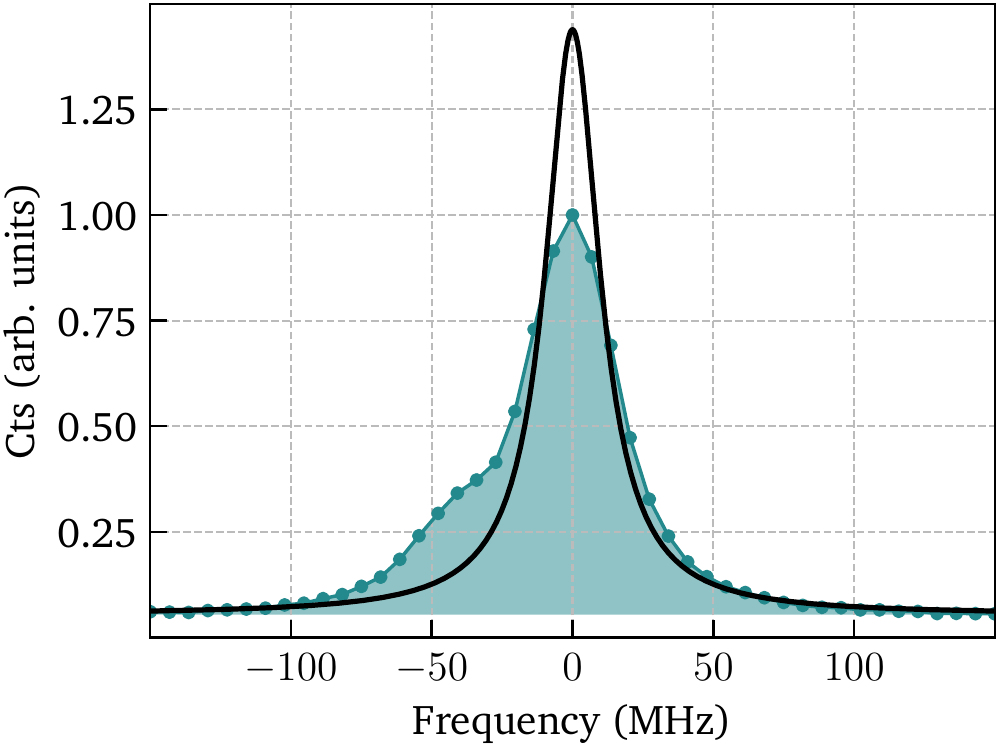}
    \caption{A generic resonance recorded at the COALA beam line. The asymmetry is caused by a drop in the source potential and a second velocity class of ions is generated at a second hotspot \cite{Konig.2020}. In black, the calculated equivalent Lorentzian signal with height $S_\textrm{L}$ and the natural linewidth is plotted, which would have been observed without broadening effects. The area under both curves is identical.}
    \label{fig:ssummed}
\end{figure}
Optical resonances at resonance frequency $\nu_0$ between bound states in the absence of inhomogeneous broadening are of Lorentzian shape, with
    \begin{align}
        \dot{R}_\textrm{L} (\nu) = \frac{A_\textrm{L}}{\pi}\frac{\frac{1}{2} \Gamma}{\left(\nu-\nu_0\right)^2+\left(\frac{1}{2} \Gamma\right)^2}~.
    \end{align}
In the case of the natural linewidth as the dominant homogeneous contribution and a transition from the ground state, their width $\Gamma$ (for $\Gamma^2/\nu_0^2 \ll 1$) is given by 
    \begin{align}
        \Gamma_\textrm{nat} = \frac{1}{2 \pi \tau}
    \end{align} 
with the lifetime $\tau$ of the upper state and denoting the full width at half maximum (FWHM) of the distribution. In real experiments, Lorentzian shape is only observed in the very special case of ultra-high resolution measurements on very cold ensembles $(T \rightarrow 0\,K)$. Usually, one or several broadening mechanisms induce an additional linewidth. Most prominently, the velocity distribution of the particles from the source contributes as Doppler broadening. Sometimes, such broadening mechanisms can be explicitly expressed mathematically and convoluted with the Lorentzian shape. This is particularly true for the Doppler broadening and the associated Gaussian distribution of velocities. It is regularly taken into account by applying the Voigt profile which is a convolution of the Gaussian frequency distribution
    \begin{align}
        \dot{R}_\textrm{G}(\nu) = \frac{A_\textrm{G}}{\sigma \sqrt{2\pi}} 
        e^{-(\nu-\nu_0)^2/2\sigma^2}~,
    \end{align}    
and the Lorentzian distribution, according to
    \begin{align}
        \dot{R}_\textrm{V}(\nu) = \int\limits_{-\infty}^{+\infty} \dot{R}_\textrm{G}(\nu') \, \dot{R}_\textrm{L}(\nu-\nu')\, \mathrm{d}t~.
    \end{align}
To compare the efficiencies of different photon detection units, it is necessary to normalize the observed signal rates, since the height (the maximum) of all distributions scale with their width. An efficient method is to integrate the total measured signal strength 
    \begin{align}
        S_{\textrm{tot}} = \int\limits_{-\infty}^{+\infty} \dot{R}(\nu)~\mathrm{d}\nu
    \end{align} 
which is equivalent to adding the number of photons in each recorded bin minus the fitted background. To give this value a physical meaning, it can be scaled with the Lorentzian width to
    \begin{align} \label{eq:s_l}
        S_{\mathrm{L}} = \frac{2}{\pi} \frac{S_{\textrm{tot}}}{\Gamma_\mathrm{nat}}
    \end{align}
using the natural linewidth (in units of the bin width) of the investigated transition. The obtained value $S_{\textrm{L}}$ is the height of the Lorentz distribution in the idealized case of a mono-energetic collimated beam without any additional broadening mechanisms. In the desirable scenario that every particle interacts with the laser beam not more than once, this allows to compare transitions that have different shapes and widths. As an example, Fig.~\ref{fig:ssummed} shows a generic spectrum recorded experimentally at the COALA setup which shows asymmetry due to an inhomogeneous source starting potential. The area under the blue curves equals the area under the black curve which has the signal height $S_\textrm{L}$ that would occur if the lineshape was not distorted.


\subsubsection{Particle efficiency}
The particle efficiency $\eta_\textrm{part}$ gives the number of ions or atoms for one detected photon, refering to the portion of the beam that is prepared in the right state and passing the detection region. It can be calculated by dividing the integrated particle beam current per bin divided by the summed signal height $S_{\textrm{L}}$. This of course is the extrapolated value if only the natural line width and no other broadening mechanisms were observed, and allows to compare different experimental conditions. The upper limit is given by the total detection efficiency $\epsilon_\textrm{tot}$ of the fluorescence detection system. 

Besides the difficulty of a realistic estimate of $\epsilon_\textrm{tot}$, a large discrepancy of $\eta_\textrm{part}$ might indicate non-optimal alignment or insufficient laser power. Thus, it is a critical parameter to investigate and optimize the experimental conditions prior to any experiment. 

\subsubsection{Signal-to-Noise ratio}
The signal-to-noise ratio or \texttt{SNR} is defined as
\begin{align}
    \texttt{SNR} = \frac{S_{\textrm{L}}}{\sqrt{D}}
\end{align}
where $D$ is the background level. When a sufficient number of events are recorded, the number of events in each bin are distributed Poisson-like, and the square root is a good aproximation for the statistical uncertainty. The \texttt{SNR} then gives the ratio between the normalized signal height and the statistical fluctuations of the recorded dataset. It is useful to determine the quality of a given dataset, but it is a complex parameter when comparing two separate experimental setups. A better indicator for a good detection system is the time it takes to generate a spectrum with a certain \texttt{SNR}; However, even then it does depend on beam current, laser power, background rate and measurement time and is strongly correlated with alignment and beam overlap parameters which are difficult to quantify. For this reason, datasets from different setups have to be compared very carefully regarding their \texttt{SNR}.

\subsection{COALA} \label{sec:coala}
\begin{table}[tb]
    \centering
    \ra{1.3}
    \caption{The different compound parabolic concentrator designs were mounted on the COALA beam line, and signals were recorded under constant experimental conditions. The signal and background light rates are then normalized to equalize beam current, laser power, lineshape and 1\,s measurement time. The QPC20 achieves the highest signal-to-noise ratio (\texttt{SNR}) in the given experiment but accepts the least signal photons. \textcolor{blue}{Provide a short description of each column head.}}
    \label{tab:cpc_exp}
    \begin{tabularx}{0.65\textwidth}{@{}lXlrll@{}}
    \tls
    \toprule\toprule
                &&  $S_\textrm{L}$           & $D_\textrm{static}$    & \texttt{SNR}  &  rel. \texttt{SNR}    \\
                
                &&          &   kHz/100$\mu$W  &  &     \\
    \midrule
    QPC20       && 123.4                     & 2.7                     & 2374.4             & 1.46                                               \\
    QPC30       && 131.2                    & 4.0                     & 2074.4              & 1.28                                               \\
    DC       && 137.6                    & 7.2                    & 1621.3              & 1.00                                               \\
    \bottomrule\bottomrule
    \end{tabularx}
\end{table}
The COALA beam line is a collinear laser spectroscopy setup located at TU Darmstadt \cite{Konig.2020}. It is used for a variety of applications, such as high voltage metrology \cite{Kramer.2018}, high-precision tests of atomic theory \cite{Imgram.2019} and reference measurements on stable isotopes with high accuracy. A fluorescence detection region of the type discussed here is employed using polished aluminium mirrors to provide the highest flexibility in wavelength. 

A series of experiments was conducted to evaluate the performance of the fluorescence detection system and especially the different concentrator systems. For this, a beam of calcium ions was generated by surface ionization. No contaminating elements are found in the beam, and thus its isotopic composition conforms with the natural abundance of calcium, with 97\,\% being $^{40}$Ca which is tested. This beam is then superimposed with a laser beam which was tuned to the the Doppler-shifted $4s\,^2\mathrm{S}_{1/2} \rightarrow 4s\,^2\mathrm{P}_{1/2}$ (D1) transition wavelength of 397\,nm. The laser power and ion beam current were monitored for each single scan. We recorded resonance of the D1 line in calcium with each of the three different lightguides installed. Fluctuations in beam current and laser power are effectively canceled by taking their average. By performing measurements at different laser powers and beam currents, we also ensured that saturation was not reached, and that we can linearly interpolate our data for normalization. Thus, such a dataset allows us to compare the three different lightguides under similar conditions.

\subsubsection{Background Rates}
The background collected by the PMTs scales approximately linear with the laser power. Thus, it is straightforward to give a background rate normalized to 100\,$\mu$W laser beam. To compare this rate to other setups, the quantum efficiency of the PMT at 397\,nm has to be taken into account, which is given as 0.27 by the manufacturer (Sens-Tech P25PC-UV). To extrapolate the background rates of the system described here to a different wavelength, also the spectral mirror reflectivity needs to be considered. With a QPC20 employed, we observe background rates of 2.7\,kHz/100$\mu$W which is significantly less than with the QPC30 (4.0\,kHz/100$\mu$W) or the PMT mounted directly to the viewport (7.2\,kHz/100$\mu$W). These numbers can also be found in Tab.~\ref{tab:cpc_exp}.

The background suppression is more pronounced than expected from the simulations. However, a simple model was choosen in the simulation where the background light is solely diffracted at the entrance aperture of the oval mirror. In reality, it will be scattered as well on apertures which are further away from the mirror system. This light will diverge from the laser beam, and enter the oval mirror with shallow angles that will be filtered out much more efficiently by more restrictive concentrators.

\subsubsection{Particle Efficiency}
The collected data also allows to extract values for the signal collection efficiency. Datasets were recorded for each of the three setups with similar laser power (216\,$\mu$W) at different beam currents. Normalizing to the number of ions, the data presented in Tab.~\ref{tab:cpc_exp} shows that the DC setup performs best in terms of particle efficiency $\eta_\textrm{part}$, closely followed by the QPC30 and the QPC20. This follows the expectations from the simulations which did not predict a significant difference in light collection efficiency between the three systems. The fact that the QPC30 slightly outperforms the QPC20 is a hint that the ion beam diameter is spread out which lets the more restrictive angular cutoff crop more signal photons. 

The particle efficiency $\eta_\textrm{part}$ ranges around 5000 calcium ions per photon at 216\,$\mu$W, which is worse than typical values achieved in online spectroscopy. We attribute this to the nonexisting beam cooling stage at COALA. Since spectroscopy is done on a beam directly after the thermal source, a relatively large portion of the ion beam is not in overlap with the laser beam. To get a comparative value, an experiment on a cooled, bunched $^{36}$Ca beam at BECOLA with a copy of the fluorescence detection region is analyzed in the next subchapter.

\subsubsection{Signal-to-Noise Ratio}
The \texttt{SNR} that is generated within a certain time is a decisive parameter for any experiment which relies on beam that is not infinitely available. Although the QPC20 is least efficient in collecting signal light, it is most efficient in discriminating stray light. In combination, the generated \texttt{SNR}, listed in Tab.~\ref{tab:cpc_exp} is better than the two other setups which are compared. Under the given experimental conditions, it is advisable to use a QPC20 to build up the best signal (in terms of \texttt{SNR}) in the shortest time.

There are, however, many experiments where background light is not just stray light generated by apertures, but the beam itself emits photons off-resonantly, independently of the applied laser light. A typical example for this is an atom beam being produced in a alkali-vapor filled charge exchange cell as shown in Fig.~\ref{fig:beamline}. After neutralizing, the particles send out photons while cascading to their ground state, and inelastic collisions of the beam particles with the residual gas can lead to non-resonant excitation as well. The detection system following after is not capable of differentiating between such photons that are emitted after a particle/laser interaction and random events, caused by particle/particle interaction. In some cases, especially when the beam is contaminated with other elements, color filters can help to reduce the amount of beam-induced background light. However, in all cases, a route to generate clear signals quickly is to collect as much light emitted from the beam as possible to resolve the resonance structure within the beam-related background. In highly unpure beams, which is, e.g., often obtained in an -type production process, the beam-related background outweighs the fraction of stray light that is recorded, and the net loss in signal photons for the QPC20 compared to the other two systems will result in a longer accumulation time to achieve the same \texttt{SNR}. A QPC30 or even the DC system can then provide better signals in shorter acquisition times.

\subsection{BECOLA}
\begin{figure}[ht]
    \centering
        \includegraphics[width=.65\textwidth]{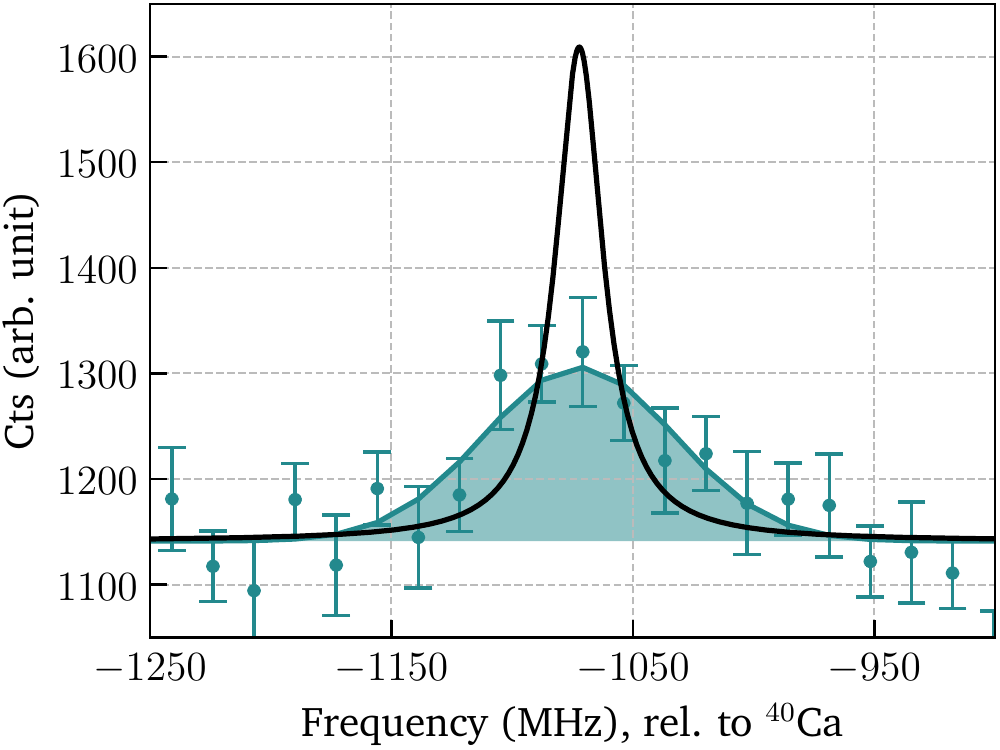}
    \caption{The resonance of $^{36}$Ca recorded at BECOLA, MSU recorded with the fluorescence detection region. The system was equipped with industrial sheet mirrors and a QPC20.}
    \label{fig:becola}
\end{figure}
The fluorescence detection region described in this chapter was used in an online experiment at BECOLA (Beam Cooling and Laser Spectroscopy) \cite{Minamisono.2013, Rossi.2014}, where the isotope shift in neutron-deficient calcium isotopes was measured to extract their nuclear charge radii~\cite{Miller.2019} and the hyperfine structure to obtain the nuclear moments \cite{Klose.2019}. $^{36}$Ca was produced at NSCL in a projectile-fragmentation reaction of $^{40}$Ca at 140\,MeV/A on a thin beryllium transmission target. The thermalized beam was transported to BECOLA where the beam was cooled and bunched in a linear buffer-gas filled Paul trap before it was superimposed with the laser beam.

In the successful experiment, the D2 line in calcium at 393\,nm was measured for the low-yield isotope $^{36}$Ca. With the cooler/buncher running in 180\,ms intervals for the $T_{1/2}=102$\,ms isotope, approximately 25 ions per second were passing through the detection region. The cooler/buncher interval was optimized to maximize the \texttt{SNR}, as shorter intervals would have lead to more ions in total but also to an increase in the total beam gate and thus the background. The laser power was set to 300\,$\mu$W. Figure\;\ref{fig:becola} shows the recorded resonance of $^{36}$Ca in one of the detection chambers, where a QPC20 is employed. The region is equipped with sheet mirrors that we tested to have $R=0.88(3)$ at 393\,nm in a simple reflection measurement. 

The area of the resonance is 1007.8\,counts, which according to Eq.~\ref{eq:s_l} corresponds to a corrected signal height of $S_\textrm{L}=467.8$. In the strongest signal bin, $\eta_\textrm{part} = 254.9$\,ions are passing through the region per photon recorded. The calculated total efficiency of the full system is $\varepsilon_\textrm{System} = 0.13$ for $R=0.88$ and calcium at 393\,nm. With the PMT quantum efficiency of 25\,\% at that wavelength, we can calculate an absolute detection efficiency of 1 in 30 photons. Thus, at a particle efficiency of $254.9$\,ions per photon, one photon is sent out inside the active area of the detection region from at least every $8^\mathrm{th}$\,ion in the beam. 

The significant difference between the particle efficiency result for the COALA beam line and the BECOLA setup can be explained by the ion beam emittance, which excels at BECOLA due to the beam cooling stage. Approximately ten times fewer ions are overlaped with the laser beam in the COALA calcium beam which is emitted from a hot carbon filament. Thus, we can conclude that despite a well-designed fluorescence detection region, a cooled low-emittance ion beam is a necessity for high-precision laser spectroscopy on low-yield isotopes.

\section{Conclusion}
In this article, we presented a fluorescence detection region for collinear laser spectroscopy which is based on curved mirrors. The in-vacuum part has an oval cross section and covers over 80\,\% of the full solid angle of an 80\,mm beam segment, forwarding photons through a vacuum viewport. Here, simulations show that stray light photons, emerging from a less localized volume, have a broader angular distribution than beam photons. This allows to passively suppress background light by employing compound parabolic concentrators, which only forward photons that enter with steep angles. A symmetric quadratic parabolic concentrator was designed and its good performance is confirmed in ray-tracing simulations and in experiment.

The combination of the oval 4$\pi$ mirror and the parabolic concentrator forwards one fifth of all beam photons to the counting photomultiplier tubes, however it requires highly-reflective mirrors for the used wavelength. While aluminium mirrors do provide sufficient reflectivity over a broad wavelength range, the detection region can be equipped with specific mirrors adapted to any given wavelength. 

First measurements with the implemented system were performed at the COALA beam line at TU Darmstadt. Here, we used a calcium beam and a 397\,nm laser and showed that the background suppression with the parabolic concentrators even exceeds conservative expectations. However, the signal photon collection efficiency is worse than in comparable setups where a better beam emittance was obtained through the beam cooling stage.

A second version of the setup is operated at the BECOLA setup at MSU. Here, the isotope shift in $^{36}$Ca was measured with ion rates of 25 per second. The data recorded in the fluorescence detection region stands out in particle efficiency, generating one signal photon per 254 ions. The only other laser spectroscopy experiment with comparably low production rates using fluorescence detection was performed at the COLLAPS setup at ISOLDE/CERN on $^{52}$Ca, with a beam of only a few hundred ions per second~\cite{GarciaRuiz.2016}, and employing a lense-based detection system.

The experiments suggest that the novel mirror-based system is well suited for a broad range of applications in laser spectroscopy experiments, including research on low-yield isotopes. In particular, the possibility to adapt the system to specific experimental conditions by employing appropriate concentrators and mirrors makes it an ideal tool in laser spectroscopy beam lines that are used for versatile applications. It also highlights that fluorescence laser spectroscopy is generally feasible even with low-yield beams by employing dedicated detection devices that are highly sensitive to the emitted photons.

\acknowledgments
We thank Michael Hammen for his early contributions to this project, and Janina Willmann who helped with the assembly and early testing. We also gratefully acknowledge the support from the BECOLA and the ALIVE team.

This project was funded by the Deutsche Forschungsgemeinschaft (DFG, German Research Foundation) - Project-ID 279384907 - SFB1245; The National Science Foundation, Grant No. PHY-15-65546; The German Federal Ministry of Education and Research (BMBF), Contract No.~05P19RDFN1. Installation of COALA and experiments were supported by the Deutsche Forschungsgemeinschaft (DFG) under Grant INST No. 163/392-1 FUGG, the Helmholtz International Center for FAIR (HIC4FAIR) and the Helmholtz Research Academy Hesse for FAIR (HFHF). K.K. and B.M. and F.S. acknowledge support from HGS-HIRE.


\bibliographystyle{altalpha}
\bibliography{biblio}

\end{document}